\documentclass[dvips]{article}
\usepackage{supertabular,lscape,epsfig}
\usepackage{amssymb}
\usepackage{amsmath}

\usepackage[polish]{babel}
\usepackage[T1]{fontenc}
\usepackage[latin2]{inputenc}
\usepackage{pslatex}

\DeclareSymbolFont{ppa}{OT1}{ppl}{m}{it}
\DeclareMathSymbol{\vv}{\mathalpha}{ppa}{'166}

\begin{document}

\newcommand{\dd}{\,{\rm d}}
\newcommand{\ie}{{\it i.e.},\,}
\newcommand{\etal}{{\it et al.\ }}
\newcommand{\eg}{{\it e.g.},\,}
\newcommand{\cf}{{\it cf.\ }}
\newcommand{\vs}{{\it vs.\ }}
\newcommand{\zdot}{\makebox[0pt][l]{.}}
\newcommand{\up}[1]{\ifmmode^{\rm #1}\else$^{\rm #1}$\fi}
\newcommand{\dn}[1]{\ifmmode_{\rm #1}\else$_{\rm #1}$\fi}
\newcommand{\upd}{\up{d}}
\newcommand{\uph}{\up{h}}
\newcommand{\upm}{\up{m}}  
\newcommand{\ups}{\up{s}}
\newcommand{\arcd}{\ifmmode^{\circ}\else$^{\circ}$\fi}
\newcommand{\arcm}{\ifmmode{'}\else$'$\fi}
\newcommand{\arcs}{\ifmmode{''}\else$''$\fi}
\newcommand{\MS}{{\rm M}\ifmmode_{\odot}\else$_{\odot}$\fi}
\newcommand{\RS}{{\rm R}\ifmmode_{\odot}\else$_{\odot}$\fi}
\newcommand{\LS}{{\rm L}\ifmmode_{\odot}\else$_{\odot}$\fi}

\newcommand{\Abstract}[2]{{\footnotesize\begin{center}ABSTRACT\end{center}
\vspace{1mm}\par#1\par   
\noindent
{~}{\it #2}}}

\newcommand{\TabCap}[2]{\begin{center}\parbox[t]{#1}{\begin{center}
  \small {\spaceskip 2pt plus 1pt minus 1pt T a b l e}
  \refstepcounter{table}\thetable \\[2mm]
  \footnotesize #2 \end{center}}\end{center}}

\newcommand{\TableSep}[2]{\begin{table}[p]\vspace{#1}
\TabCap{#2}\end{table}}

\newcommand{\FigCap}[1]{\footnotesize\par\noindent Fig.\  %
  \refstepcounter{figure}\thefigure. #1\par}

\newcommand{\TableFont}{\footnotesize}
\newcommand{\TableFontIt}{\ttit}
\newcommand{\SetTableFont}[1]{\renewcommand{\TableFont}{#1}}

\newcommand{\MakeTable}[4]{\begin{table}[htb]\TabCap{#2}{#3}
  \begin{center} \TableFont \begin{tabular}{#1} #4
  \end{tabular}\end{center}\end{table}}

\newcommand{\MakeTableSep}[4]{\begin{table}[p]\TabCap{#2}{#3}
  \begin{center} \TableFont \begin{tabular}{#1} #4
  \end{tabular}\end{center}\end{table}}

\newenvironment{references}%
{
\footnotesize \frenchspacing
\renewcommand{\thesection}{}
\renewcommand{\in}{{\rm in }}
\renewcommand{\AA}{Astron.\ Astrophys.}
\newcommand{\AAS}{Astron.~Astrophys.~Suppl.~Ser.}
\newcommand{\ApJ}{Astrophys.\ J.}
\newcommand{\ApJS}{Astrophys.\ J.~Suppl.~Ser.}
\newcommand{\ApJL}{Astrophys.\ J.~Letters}
\newcommand{\AJ}{Astron.\ J.}
\newcommand{\IBVS}{IBVS}
\newcommand{\PASP}{P.A.S.P.}
\newcommand{\Acta}{Acta Astron.}
\newcommand{\MNRAS}{MNRAS}
\renewcommand{\and}{{\rm and }}
\section{{\rm REFERENCES}}
\sloppy \hyphenpenalty10000
\begin{list}{}{\leftmargin1cm\listparindent-1cm
\itemindent\listparindent\parsep0pt\itemsep0pt}}%
{\end{list}\vspace{2mm}}
 
\def\TYLDA{~}
\newlength{\DW}
\settowidth{\DW}{0}
\newcommand{\dw}{\hspace{\DW}}

\newcommand{\refitem}[5]{\item[]{#1} #2%
\def\REFARG{#3}\ifx\REFARG\TYLDA\else, {\it#3}\fi
\def\REFARG{#4}\ifx\REFARG\TYLDA\else, {\bf#4}\fi
\def\REFARG{#5}\ifx\REFARG\TYLDA\else, {#5}\fi.}

\newcommand{\Section}[1]{\section{#1}}
\newcommand{\Subsection}[1]{\subsection{#1}}
\newcommand{\Acknow}[1]{\par\vspace{5mm}{\bf Acknowledgements.} #1}
\pagestyle{myheadings}

\newfont{\bb}{ptmbi8t at 12pt}
\newcommand{\xrule}{\rule{0pt}{2.5ex}}  
\newcommand{\xxrule}{\rule[-1.8ex]{0pt}{4.5ex}}  
\def\thefootnote{\fnsymbol{footnote}}
\begin{center}

{\Large\bf
The Optical Gravitational Lensing Experiment.
OGLE-III 
Long Term Monitoring of the Gravitational Lens QSO 2237+0305\footnote{
Based on observations obtained with the 1.3-m Warsaw
telescope at the Las Campanas Observatory of the Carnegie Institution of
Washington.}}
\vskip1.7cm
{\bf A.~~U~d~a~l~s~k~i, ~~M.\ K.~~S~z~y~m~a~{\'n}~s~k~i,
~~M.~~K~u~b~i~a~k, ~~G.~~P~i~e~t~r~z~y~\'n~s~k~i,
~~I.~~S~o~s~z~y~\'n~s~k~i, ~~K.~~\.Z~e~b~r~u~\'n, ~~O.~~S~z~e~w~c~z~y~k, 
~~\L.~~W~y~r~z~y~k~o~w~s~k~i,
~~K.~~U~l~a~c~z~y~k ~~and ~~T.~~W~i~ę~c~k~o~w~s~k~i}
\vskip5mm
{Warsaw University Observatory, Al.~Ujazdowskie~4, 00-478~Warszawa, Poland\\
e-mail:
(udalski,msz,mk,pietrzyn,soszynsk,zebrun,szewczyk,wyrzykow,kulaczyk,twieck)\\
@astrouw.edu.pl}
\end{center}
\vskip1.5cm
\Abstract{We present results of the long term monitoring of the
gravitational lens QSO 2237+0305 conducted during the OGLE survey.  Light
curves of all four components of the lens obtained during the second phase
of the OGLE project (OGLE-II; 1997--2000) are supplemented with the data
collected in the OGLE-III phase in the observing seasons
2001--2006. Calibration procedures to tie the new OGLE-III data with
already calibrated OGLE-II light curves are described. The resulting
homogeneous OGLE data set is the most extensive photometric coverage of the
gravitational lens QSO 2237+0305, spanning now one decade -- the seasons
from 1997 to 2006 -- and revealing unique microlensing activity of this
spectacular object.

All photometric data of the gravitational lens QSO 2237+0305 collected by
OGLE are available to the astronomical community from the OGLE {\sc
Internet} archive.}{}

\vskip9mm
\Section{Introduction} 
The gravitational lens QSO 2237+0305 called also Huchra's lens (Huchra
\etal 1985) or the Einstein Cross is a spectacular object consisting of
four, separated by about 1\arcs\ in the sky, images of a distant quasar
($z=1.7$) and a nearby barred spiral galaxy ($z=0.04$) that acts as a
gravitational lens. Favorable positioning of the quasar and galaxy
should potentially produce prominent microlensing effects in the light
curve of the quasar images (Schneider \etal 1988) and indeed such
effects were discovered by Irwin \etal (1989). Due to the magnitude of
microlensing variability the 2237+0305 lens became the crucial object
for testing the models of the source and mass distribution in the
lensing galaxy and our understanding of extragalactic microlensing (\eg
Wambsganss, Paczyński and Schneider 1990, Rauch and Blandford 1991,
Jaroszyński, Wambsganss and Paczyński 1992, Wyithe, Webster and Turner
2000a, Yonehara 2001, Wisotzki \etal 2003, Kochanek 2004, Jaroszyński
and Skowron 2006 and many others).

Although it was soon realized that for constraining the models long
term, well sampled monitoring of the object is necessary, the
photometric data of 2237+0305 remained sparse until the end of 1990s.
Typical light curve contained only a few epochs per season and covered
at most five observing seasons (Corrigan \etal 1991, {\O}stensen \etal
1996). One of the reasons were difficulties in measuring with reasonable
accuracy the magnitudes of lens components because of significant
blending due to their proximity in the sky and background of the lensing
galaxy. Only observations obtained at observing sites with very good
seeing  could be useful.

The situation has significantly improved when the Optical Gravitational
Lensing Experiment (OGLE) started regular monitoring of the 2237+0305
lens as a subproject of the second phase (OGLE-II) of this long term
photometric survey (Wo{\'z}niak \etal 2000a). Regular -- every few
nights -- observing pattern of 2237+0305 combined with the newly
developed algorithm of data reduction based on the image difference
technique (DIA: Alard and Lupton 1998, Alard 1999, Wo{\'z}niak 2000),
that turned out to be very successful and allowed precise determination
of magnitudes of all four components, produced unique new generation
light curves of the 2237+0305 lens images. When the data reduction
algorithm was implemented at the OGLE telescope to allow real time
photometric monitoring of 2237+0305 (1998) it became possible to start
hunting for the detection of high magnification events in the
microlensing variability (Wo{\'z}niak \etal 2000b). Such rare events, in
particular the caustic crossing which should occur every few years, are
very important for mapping the distribution of light of the lensed
quasar.

From the end of 1990s two other groups also monitored the 2237+0305 lens
simultaneously with OGLE. Complementary to OGLE  but less extensive data
sets were obtained by the CLITP project (Alcalde \etal 2002) from Canary
Islands and by Vakulik \etal (2004) and Koptelova \etal (2004) from
Majdanak Observatory.

The OGLE-II monitoring of the 2237+0305 lens ended at the end of 2000
and the OGLE-II light curves covered four observing seasons: 1997--2000.
When the OGLE-III phase (Udalski 2003) started  regular observations in
2001, the 2237+0305 lens was again included to the list of monitored
targets. In this paper we describe the collected OGLE-III data of
2237+0305 as well as the reduction and calibration procedures. The main
goal of this paper is to provide the astronomical community  with unique
homogeneous calibrated light curves of the 2237+0305 lens images
covering at present ten observing seasons from 1997 to 2006. 
\vspace*{9pt}
\Section{Observational Data} 
\vspace*{5pt}
Observations presented in this paper were collected with the 1.3-m Warsaw
telescope at the Las Campanas Observatory, Chile (operated by the Carnegie
Institution of Washington), equipped with a wide field CCD mosaic
camera. The camera consists of eight ${2048\times4096}$ pixel SITe ST002A
detectors. The pixel size of each of the detectors is 15~$\mu$m giving the
0\zdot\arcs26/pixel scale at the focus of the Warsaw telescope. Full field
of view of the camera is about ${35\arcm\times35\arcm}$. The gain of each
chip is adjusted to be about 1.3~e$^-$/ADU with the readout noise of about
6 to 9~e$^-$, depending on the chip.

The typical observing season of 2237+0305 from the Las Campanas Observatory
lasts for about 8 months, starting at the end of April and ending in mid
December each year. Because of very small separation of the 2237+0305
images strong seeing limits are imposed on the collected images. The upper
limit of acceptable seeing is 1\zdot\arcs5. Each observation of 2237+0305
consists of two exposures to have two independent measurements of
brightness for each epoch to check consistency. The position of the lens is
shifted by about $15\arcs$ between these exposures to avoid accidental
putting the object on a defect on the chip. Also repeating the exposures
ensures at least one good measurement in the case of a hit of the lens by a
cosmic ray.

The lens is always placed in the middle of chip \#3 of the OGLE mosaic
camera. Observations are obtained through the {\it V}-band filter and
the exposure time is 360 seconds. 2237+0305 is monitored every 3--5
nights, weather permitting. If the brightness of any component starts
rising fast, observations are made more frequently, sometimes even every
night. The first two observing seasons -- 2001 and 2002 -- contain fewer
observations. Although the 2237+0305 lens was included to the OGLE
targets from the very beginning, the OGLE-III project started regular
observations in mid June 2001 -- well in the 2237+0305 observing season.
In the 2002 season the 2237+0305 lens was observed rarely due to some
technical problems. 
\vspace*{9pt}
\Section{Data Reductions} 
\vspace*{5pt}
Collected images of 2237+0305 are de-biased and flatfielded at the
telescope by the regular OGLE-III pipeline (Udalski 2003). Then they are
fed into the main reduction pipeline which includes a separate procedure
for deriving photometry of 2237+0305. It is based on the image difference
technique, DIA, and is very similar to the OGLE-III standard reductions of
relatively empty fields in the LMC and SMC halos.  In the case of 2237+0305
lens only a $2048\times2048$ pixel subframe centered on the lens is
photometrically reduced. Although the 2237+0305 field is very empty, such a
subframe contains about 50 stars -- sufficiently enough to derive accurate
transformations between reduced and reference images: astrometric
transformation for resampling the reduced image to the common $(X,Y)$ grid
and PSF transformation for image subtraction. The reference image of
2237+0305 is constructed from 18 individual, best seeing images. A subframe
of this image showing the lensing galaxy and four quasar images is shown in
Fig.~1.
\begin{figure}[htb]
\centerline{
\includegraphics[width=5.7cm, angle=90]{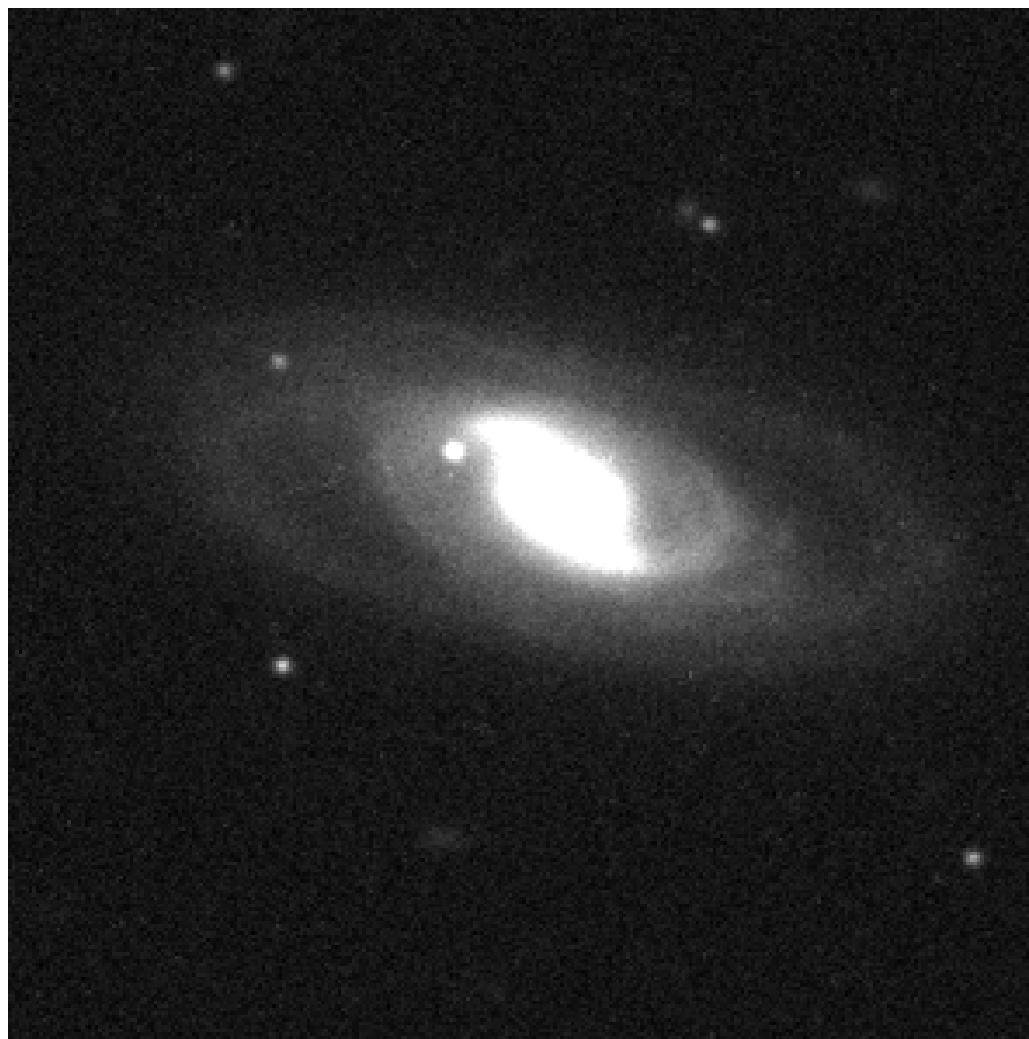}
\includegraphics[width=5.61cm]{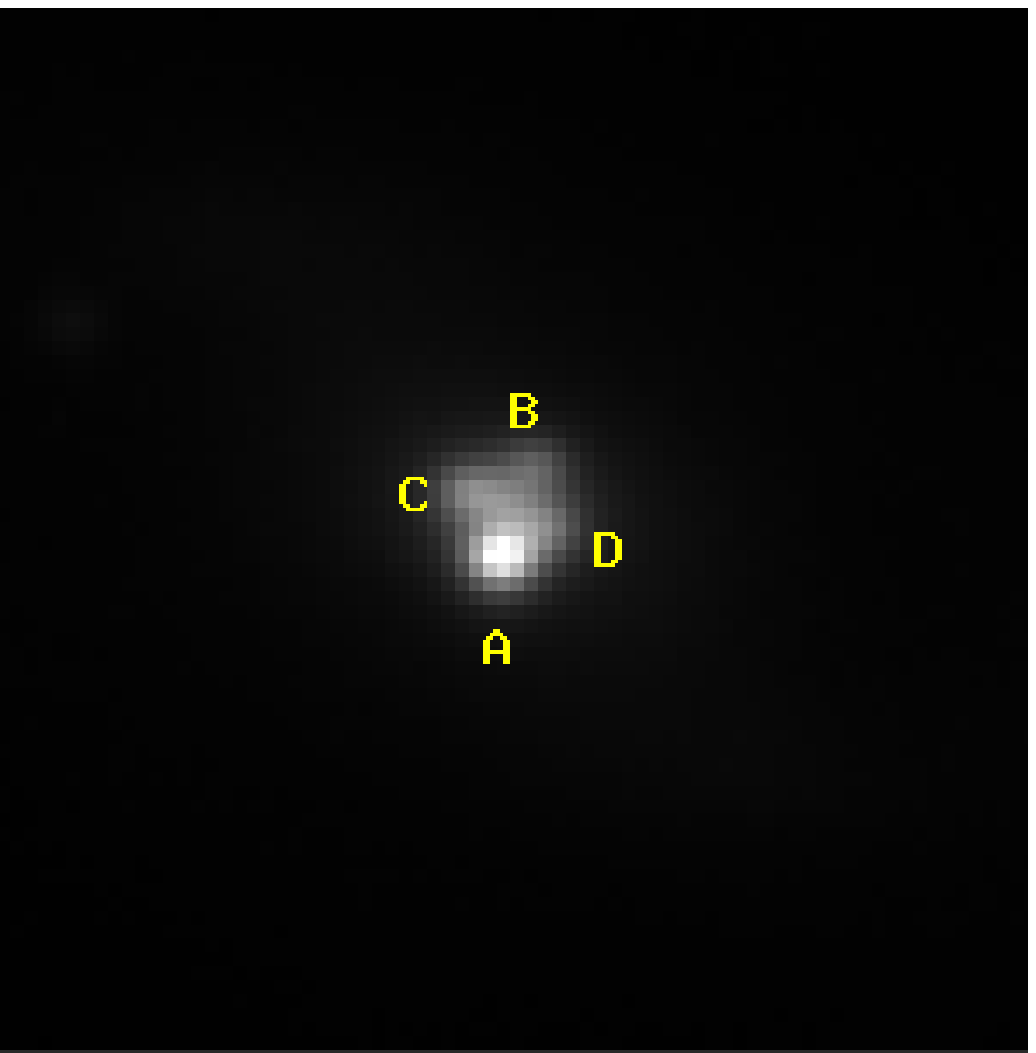}
}
\vskip 10pt
\FigCap{Lensing galaxy and four images of the gravitational lens QSO
2237+0305. Right subframe shows cental part of the galaxy zoomed four
times. North is up, East to the left.}
\end{figure}

After image subtraction the remaining signal contains information on the
variability of components. We measure these differential fluxes with PSF
photometry fitting simultaneously four profiles centered at fixed
position of each component derived from the HST image, similar as in
Wo{\'z}niak \etal (2000a). Photometry is also obtained at the position of a
few reference stars. To obtain the absolute flux of each measured object
the difference signal must be added to the reference image flux. For
constant stars this flux is measured with PSF fitting program ({\sc
DoPHOT}) and converted to the DIA flux scale. However, the 2237+0305
components cannot be reliably measured in the standard way in the
reference image due to large and  non-uniform background of the lensing
galaxy. Therefore the reference fluxes of the 2237+0305 components must
have been derived in other way (Section~4).

Finally, the measured absolute fluxes of our objects were converted to
magnitudes and shifted to the calibrated magnitude scale by adding a
zero point based on the known {\it V}-band magnitudes of reference stars
and stored in the standard OGLE database of the 2237+0305 field. 

\Section{Calibration and Tests of the Photometry}
Although during the observing seasons 2001--2003 observations of 2237+0305
were regularly carried out, the collected images had to wait for
photometric reductions until the significant sample of good seeing frames,
potential components of the reference image, was secured. It is crucial to
achieve the best photometric accuracy to stack about 10--20 best seeing
images when building the reference image. After 2003 season the dataset was
complete enough that a good quality reference image could be made and the
DIA photometry of already collected data could be derived. Since the
beginning of 2004 season the real time data system was installed at the
telescope for on-line reductions of the 2237+0305 observations.

While it was easy to tie the magnitude scale of the OGLE-III photometry of
stars in the field of 2237+0305 to the standard photometry scale obtained
during the OGLE-II phase based on non-variable comparison stars (Wo{\'z}niak
\etal 2000a), the determination of magnitudes of the lens components
required information on their reference image fluxes -- the values unknown
{\it a priori} and very difficult to measure because of small angular
components separation and large background of the overlapped lensing
galaxy. The DIA photometry only provides precise difference fluxes between
measured and reference images.

Contrary to non-variable stars the reference fluxes of the lens components
were certainly different than those used during OGLE-II monitoring (after
matching both photometric scales). Because the four images of 2237+0305 are
highly variable, their flux values in the reference image constructed from
2001--2003 frames must have been different than those in the OGLE-II
reference image from 1990s. On the other hand, the lens components fluxes
in OGLE-II reference image and, thus, the entire OGLE-II photometry were
calibrated to the standard {\it V}-band scale by Wo{\'z}niak \etal
(2000a). Unfortunately, there was more than half a year gap between the
last OGLE-II and first OGLE-III observations so it was not possible to
compare almost simultaneous observations from both OGLE observing set-ups
and to tie both photometries of the lens.  Therefore, as the first
approximation we extrapolated the observed OGLE-II light curves of each
lens component to the epoch of the initial OGLE-III observations of the
2237+0305 lens. Then, we used these extrapolated, calibrated magnitudes to
derive the necessary OGLE-III reference image fluxes of each component that
produced identical as OGLE-II magnitudes for the initial OGLE-III
images. However, such an approach was only a crude approximation of the
OGLE-III magnitude scale of lens images and its accuracy highly depended on
the unknown variability of the lens components during the eight months gap
in observations between OGLE-II and OGLE-III phases. Nevertheless, we
decided to make our photometry publicly available at that moment. The
general shape of the light curve was in any case preserved so the main goal
of the real time monitoring system -- prompt detection of rapid brightness
changes could have been achieved. Up to the end of the 2006 observing
season the OGLE-III data posted on the 2237+0305 OGLE WWW page used this
preliminary calibration.

It is obvious that the application of the unique OGLE observations of
2237+ 0305 to any scientific project requires much more precise
calibration. In particular, it is very important that both OGLE-II and
OGLE-III data sets are accurately tied photometrically to each other making
one homogeneous data set covering now ten seasons. Therefore, we applied
the following procedure. First, we assumed that the calibration of OGLE-II
magnitudes of the lens components (Wo{\'z}niak \etal 2000a) is correct and we
did not attempt to independently recalibrate these zero points. To tie the
OGLE-II and OGLE-III 2237+0305 photometries we decided to reduce a sample
of the best seeing OGLE-II images of 2237+0305 with the OGLE-III data
pipeline, that is with the OGLE-III reference image with provisional
approximate calibration of the lens fluxes. Because the calibrated
magnitudes of components are known for this sample from OGLE-II dataset,
the results of new reductions of the sample with OGLE-III reference image
can be used to readjust OGLE-III reference fluxes of the lens and to tie
both datasets.

It is also worth mentioning that such an approach simultaneously provided a
test of reliability of OGLE reductions of 2237+0305 and observed features
in the light curves. Because the new reductions of a sample of OGLE-II
images must have been done with completely different OGLE-III reference
image, different version of the DIA software and required special
preparation of the OGLE-II frames, the consistency of the new reductions
with the original light curves would provide strong argument that the
observed light variations are real, not of instrumental/reduction origin.

\begin{figure}[htb]
\centerline{\includegraphics[width=11cm, bb=10 40 510 710]{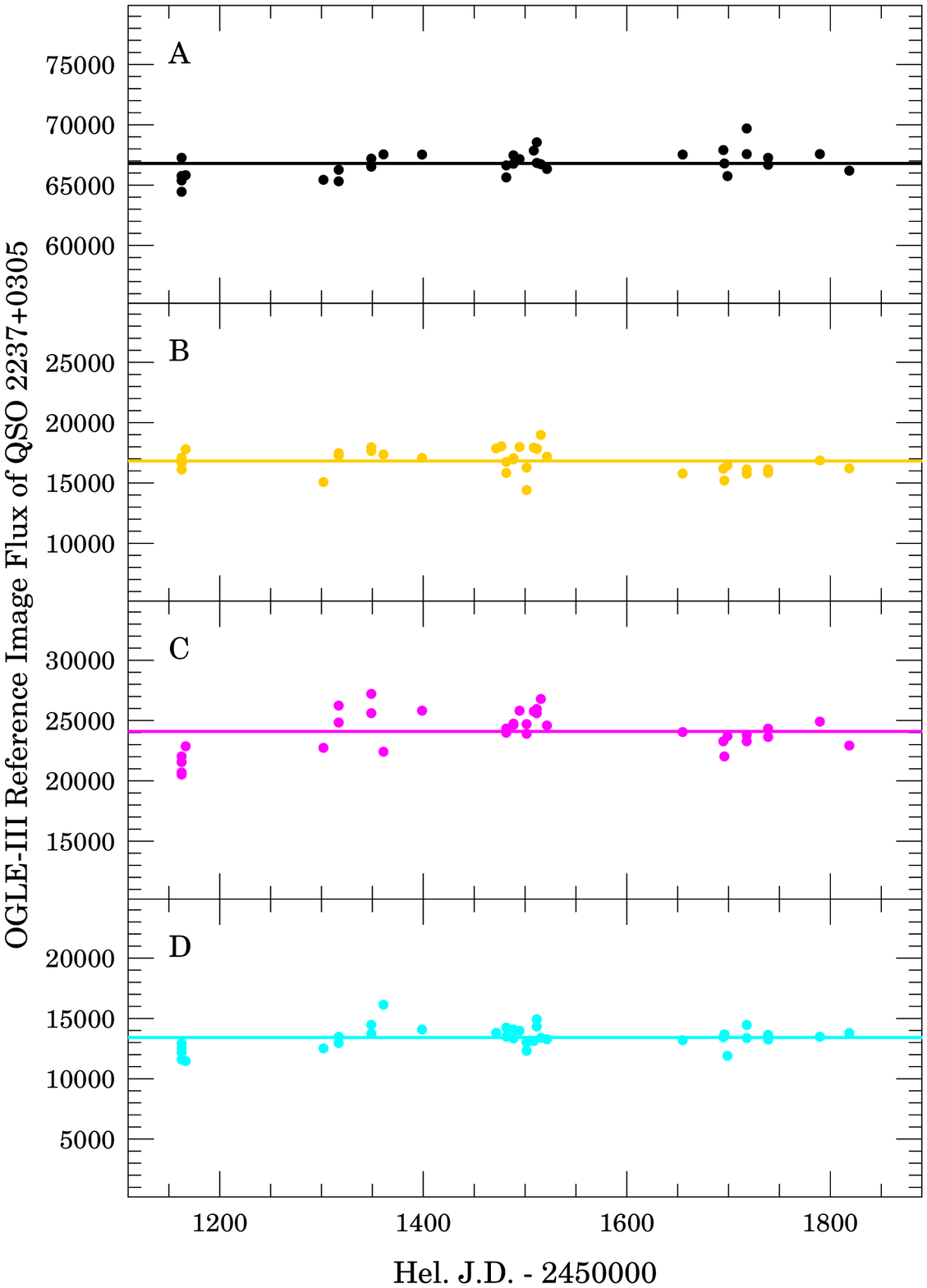}}
\FigCap{OGLE-III reference image flux of each of the components of 2237+0305.}
\end{figure}

We selected a sample of 36 best seeing frames of 2237+0305 collected during
OGLE-II after HJD=2\,451\,110. The position of 2237+0305 on earlier images
was shifted so that the OGLE-II field would only partially overlap with the
OGLE-III reference image lowering the already small number of field stars
necessary for the DIA image transformations. The selected OGLE-II images
required special preparation before running through the OGLE-III data
pipeline. First they had to be cleaned from cosmic ray hits and then
rotated by $\approx90\arcd$, flipped and resampled to the OGLE-III image
scale (from 0\zdot\arcs42/pixel in OGLE-II to 0\zdot\arcs26/pixel in
OGLE-III). Resampling was performed using flux conserving interpolation
with bicubic splines. Due to the necessity of resampling of the OGLE-II
images to the finer resolution grid we limited our calibrating OGLE-II
images only to those with the best resolution (seeing $<1\zdot\arcs25$).

\MakeTable{cl}{12.5cm}{Reference image fluxes of the 2237+0305 components}
{\hline
\noalign{\vskip3pt}
Image & \multicolumn{1}{c}{Flux}\\
\hline
\noalign{\vskip3pt}
A  &  $66794\pm1043$\\
B  &  $16828\pm978$\\
C  &  $24099\pm1624$\\
D  &  $13417\pm921$\\
\noalign{\vskip3pt}
\hline}
After running the OGLE-II images through the OGLE-III pipeline we compared
the OGLE-II calibrated magnitudes (Wo{\'z}niak {\etal 2000a) with new OGLE-III
pipeline photometry and for each lens component and each OGLE-II image we
derived the corrected OGLE-III reference image flux.  Fig.~2 presents the
results as a function of time. It can be seen that the results are very
consistent -- the new OGLE-III reference image flux of each lens component
is, within the errors, constant over the time.  The average values of
these, now calibrated to OGLE-II scale, fluxes are listed in
Table~1. Accuracy of their determination varies from about 2\% for the
brightest image A to about 7\% for the faintest image D. The magnitude of
each component can be derived from Eq.~(1). This is a prescription how to
rescale OGLE-III data if the fluxes from Table~1 (that is the Wo{\'z}niak \etal
(2000a) calibration) were again recalibrated.
$$V=-2.5\cdot\log({\rm Difference\_flux+Reference\_flux})+29.20.\eqno(1)$$

Fig.~3 shows the original OGLE-II light curves of 2237+0305 components
(small dots) with superimposed  photometry of our sample of OGLE-II best
seeing images processed with the OGLE-III pipeline (large dots). It is
clear that the agreement and consistency of both light curves is very
good indicating that with the OGLE-III reference image fluxes from
Table~1 the OGLE-III photometry of 2237+0305 is now in the calibrated
OGLE-II scale. Also, remarkable agreement of the shape of light curves
from  both completely independent reductions of OGLE-II images strongly
indicates that the measured variations of brightness of 2237+0305 are
real.

\begin{figure}[p]
\centerline{\includegraphics[width=11.3cm, bb=10 40 510 710]{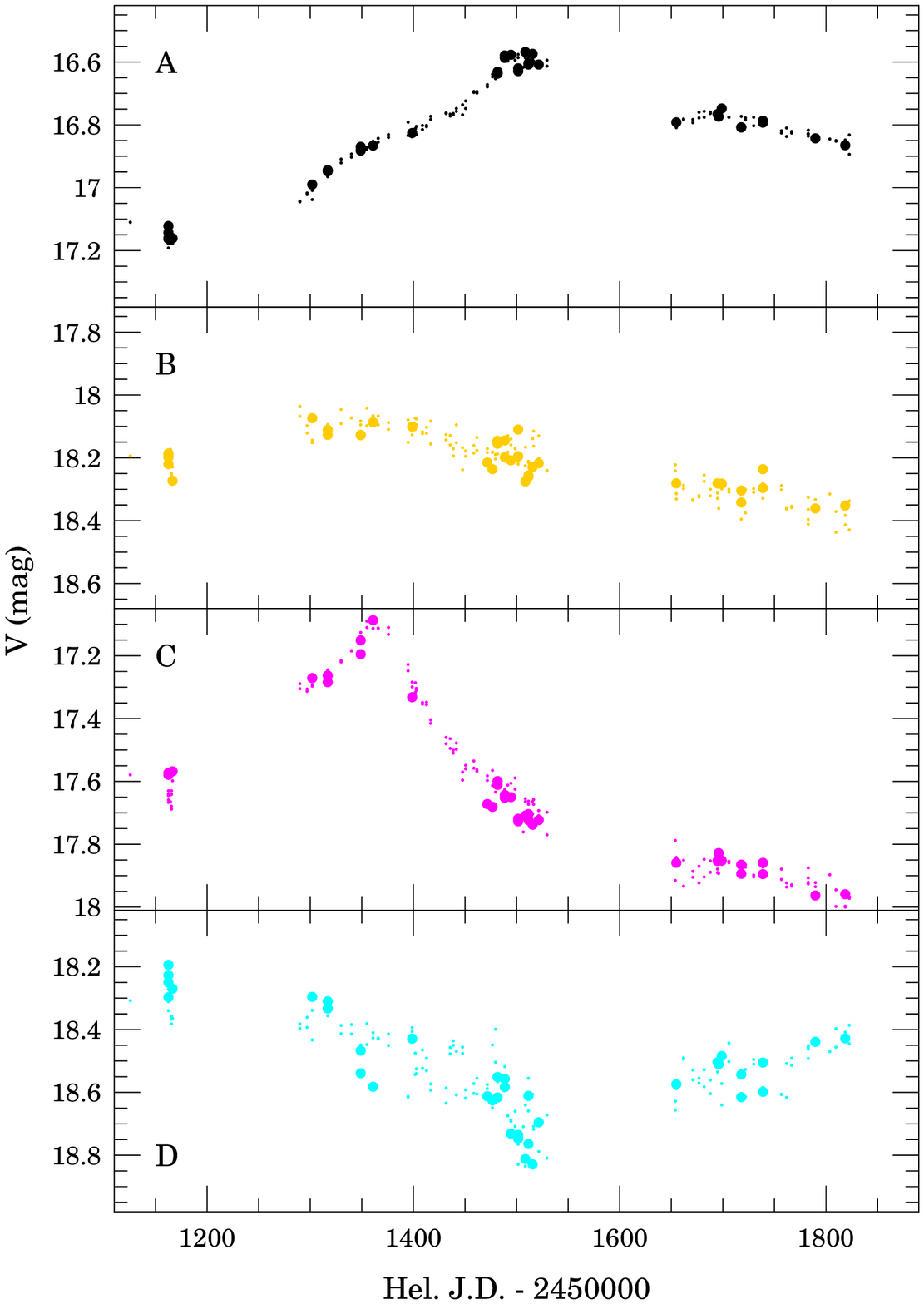}}
\FigCap{OGLE-II light curves of all images of 2237+0305 (small dots)
with photometry of the best seeing OGLE-II images reduced {\it via} the
OGLE-III data pipeline with reference image fluxes from Table~1 (large
dots).}

\vskip5mm
\normalsize
Finally, as the last step we converted the provisionally calibrated
photometry of 2237+0305 components collected during the entire OGLE-III
observing period (2001--2006) to the calibrated OGLE-II scale recalculating
the magnitudes using the calibrated reference image fluxes from Table~1.
In this way we obtained one homogeneous dataset covering the years
1997--2006. Table~2 presents calibrated photometry of 2237+0305 images and
two reference stars (Wo{\'z}niak \etal 2000a) collected during OGLE-III.  Only
several first observations are listed as an example -- the full Table is
available electronically from the OGLE WWW archive -- see Section~6. It is
regularly updated during the observing season when new observations are
collected. Also OGLE-II photometry is available from this site.
\end{figure}

\renewcommand{\TableFont}{\scriptsize}
\MakeTable{c@{\hspace{5pt}}c@{\hspace{5pt}}c@{\hspace{5pt}}c
@{\hspace{5pt}}c@{\hspace{5pt}}c@{\hspace{5pt}}c}{12.5cm}
{OGLE-III photometry  of the QSO 2237+0305 gravitational lens (sample)}
{\hline
\noalign{\vskip3pt}
HJD & A & B & C & D & C1 & C2 \\
$-2\,450\,000$ & [mag] & [mag] & [mag] & [mag] & [mag] & [mag] \\
\noalign{\vskip3pt}
\hline
\noalign{\vskip3pt}
2085.89230 & $17.028{\pm}0.007$ & $18.588{\pm}0.021$ & $18.051{\pm}0.014$ & $18.747{\pm}0.020$ & $17.412{\pm}0.005$ & $18.160{\pm}0.008$ \\ 
2085.89738 & $17.013{\pm}0.008$ & $18.538{\pm}0.022$ & $18.053{\pm}0.015$ & $18.884{\pm}0.024$ & $17.403{\pm}0.005$ & $18.170{\pm}0.008$ \\
2091.89517 & $17.024{\pm}0.005$ & $18.536{\pm}0.014$ & $18.062{\pm}0.010$ & $18.777{\pm}0.014$ & $17.409{\pm}0.004$ & $18.181{\pm}0.006$ \\
2091.90025 & $17.029{\pm}0.006$ & $18.569{\pm}0.015$ & $18.079{\pm}0.011$ & $18.832{\pm}0.015$ & $17.407{\pm}0.004$ & $18.177{\pm}0.007$ \\
2103.81800 & $17.052{\pm}0.005$ & $18.529{\pm}0.013$ & $18.061{\pm}0.010$ & $18.764{\pm}0.012$ & $17.400{\pm}0.004$ & $18.155{\pm}0.007$ \\
2103.82308 & $17.057{\pm}0.005$ & $18.479{\pm}0.012$ & $18.061{\pm}0.009$ & $18.784{\pm}0.011$ & $17.395{\pm}0.004$ & $18.156{\pm}0.007$ \\
\noalign{\vskip3pt}
\hline}

\Section{Discussion}
Calibration of the OGLE-III observations of 2237+0305 described in the
previous Section indicates that the provisional calibration presented in
OGLE WWW archive in years 2004--2006 was relatively accurate to several
percent for images A, B and C. The largest discrepancy occurred for
image D that showed a steep rise of brightness in the end of 2000
OGLE-II season leading to overestimated extrapolated magnitude for the
beginning of the 2001 OGLE-III season. Fortunately this image was
relatively quiet during the further OGLE-III monitoring.

The test of accuracy of the method of reduction of 2237+0305 is
reassuring that the features observed in the derived light curves are
real. Fig.~3 clearly shows that in spite of different reference images,
different pixel sampling of observations, orientation of the object on
the chip, different reduction software etc., the light curve features
remain practically identical. Thus, one can be confident that the entire
light curve represents the real variation of brightness of the 2237+0305
images. Similar shape of the light curves of 2237+0305 images was also
obtained by Alcalde \etal (2002) and Koptelova \etal (2004) in the
overlapping period of observations even though these observations were
obtained through a different filter.

Fig.~4 presents the entire light curve of all four components of
2237+0305 covering one decade of OGLE observations: 1997--2006. This is
the most extensive existing photometric dataset of 2237+0305 showing
complex long term variability of this unique object. The light curves of
each image reveal large brightness variations in the time scale of
months and years. The largest brightenings are certainly caused by
microlensing activity. However, part of the variability can probably be
also attributed to the long time variation of the quasar brightness. The
time delay of 2237+0305 is believed to be very short (Wambsganss and
Paczyński 1994, Dai \etal 2003, Kopteleva, Oknyanskij and Shimanovskaya
2006), so the correlated variations present sometimes in all images
suggest source brightness changes.

Practically each of the images underwent a major episode of brightening
during the last decade. During the OGLE-II phase the most spectacular ones
included $\approx0.8$~mag brightening of the image A with the peak in 1999
(${\rm HJD'}\equiv{\rm HJD-2\,450\,000}\approx 1500$) and 
$\approx1.2$~mag brightening of
the image C (peaking at ${\rm HJD'}\approx 1350$) that for a while became
the second brightest one. Both these episodes were interpreted either as a
possible caustic crossing or cusp approach and thoroughly analyzed (\eg
Wyithe, Webster and Turner 2000b, Gil-Merino \etal 2006).  The remaining
images B and D, while showed considerable variability, varied in longer
time scales.

\begin{landscape}
\begin{figure}[p]
\hglue-5.7cm
\vglue-6.5cm
\includegraphics[width=17.7cm, angle=90]{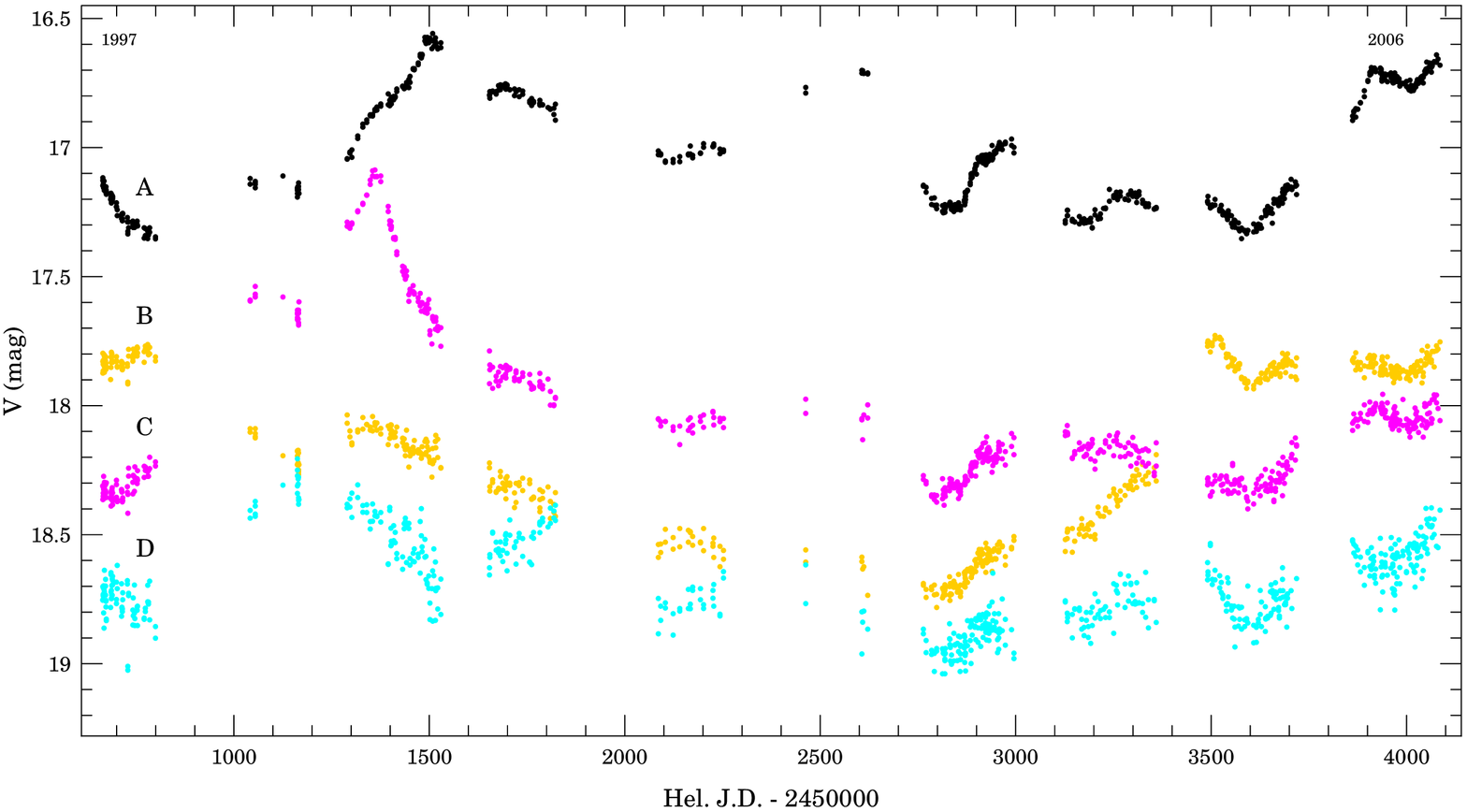}
\vskip 16pt
\FigCap{OGLE light curve of the gravitational lens QSO 2237+0305
covering ten observing seasons 1997--2006.}
\end{figure}
\end{landscape}

Images B, C and D were relatively quiet during the first years of OGLE-III
monitoring while image A showed a $\approx 0.5$~mag variability in long
time scale. The most spectacular changes occurred during the last three
observing seasons. First, the image B started brightening again and after
the rise by about 1~mag, became again the second brightest one (maximum at
${\rm HJD'}\approx3500$). On the other hand the image A began to increase
its brightness in the middle of the 2005 season and brightened continuously
in the first part of the 2006 season. The rising was so rapid that at some
moment it suggested an approach to the caustic. However, it slowed down in
the first months of the 2006 season leading to a peak at ${\rm
HJD'}\approx3900$. Nevertheless, the brightness remained at similar high
level during the next months. In the second part of the 2006 season image A
started brightening again. End of the 2006 season prevented further
observations of this extremely interestingly developing episode. It will be,
then, crucial to collect the first observations of 2237+0305 as soon as
possible in April 2007.

Although the variability of all images of 2237+0305 during the last
decade showed remarkable events, there was no clear and unambiguous
evidence of the caustic crossing in any of the images.  Theoretical
predictions indicate that in similar time scales of the OGLE monitoring 
one could expect occurrence of at least one such event (Wyithe, Webster
and Turner 2000c). Thus, the hunt for the caustic crossing in 2237+0305
(Wo{\'z}niak \etal 2000b) still remains the most important goal of the
further monitoring of 2237+0305. The OGLE survey plans to continue
similar coverage of this gravitational lens during the next observing
seasons.

\Section{Data Availability}
The photometric data of OGLE-III monitoring of 2237+0305 are available
in the electronic form from the OGLE archive:

\centerline{\it http://ogle.astrouw.edu.pl/} 
\centerline{\it http://ogle.astrouw.edu.pl/ogle3/huchra.html} 
\centerline{\it ftp://ftp.astrouw.edu.pl/ogle/ogle3/huchra}

During the observing season (April--December) the light curves are
updated regularly after each observation providing real time monitoring
of the gravitational lens QSO 2237+0305.

\Acknow{The paper was partly supported by the Polish MNiSW DST grant
to Warsaw University Observatory. Partial support to the OGLE project was
provided with the NSF grant AST-0607070 and NASA grant NNG06 GE27G to
B.~Paczy\'nski.}

\end{document}